\documentclass[%
reprint,
superscriptaddress,
amsmath,amssymb,
aps,
]{revtex4-1}

\usepackage{gensymb}
\usepackage{graphicx}
\usepackage{dcolumn}
\usepackage{bm}
\usepackage{color}
\usepackage{braket}
\usepackage[normalem]{ulem}
\usepackage{caption}
\usepackage{subcaption}

\newcommand\ai{{\it ab initio}}
\newcommand\ea{{\it et al.}}

\newcommand\ie{{\it i.e.}}

\newcommand\Tc{T$_\text{C}$}
\newcommand\Te{T$_\text{e}$}

\begin{document}
	
	\preprint{APS/123-QED}
	
	\title{Uncovering the role of the light-induced reduction of the inter-atomic exchange in the ultrafast demagnetization of Fe, Co, and Ni}
	
	\author{Philippe Scheid}
	\email{philippe.scheid@univ-lorraine.fr}
 	\affiliation{
	Department of Physics and Astronomy, Uppsala University, Box 516, SE-75120 Uppsala, Sweden
	}
	\affiliation{
   Universit\'e de Lorraine, LPCT, CNRS, UMR 7019, BP 70239,  54506 Vandoeuvre-l\`es-Nancy Cedex, France
	}
	\affiliation{
	Universit\'e de Lorraine, IJL, CNRS, UMR 7198, BP 70239, 54000 Nancy Cedex, France
	}

 	\author{Julius Hohlfeld}
	\affiliation{
	Universit\'e de Lorraine, IJL, CNRS, UMR 7198, BP 70239, 54000 Nancy Cedex, France
	}
	\author{Gregory Malinowski}
	\affiliation{
	Universit\'e de Lorraine, IJL, CNRS, UMR 7198, BP 70239, 54000 Nancy Cedex, France
	}

 	\author{Michel Hehn}
	\affiliation{
	Universit\'e de Lorraine, IJL, CNRS, UMR 7198, BP 70239, 54000 Nancy Cedex, France
	}

 	\author{Wei Zhang}
	\affiliation{
	Universit\'e de Lorraine, IJL, CNRS, UMR 7198, BP 70239, 54000 Nancy Cedex, France
	}

  	\author{Anders Bergman}
	\affiliation{
	Department of Physics and Astronomy, Uppsala University, Box 516, SE-75120 Uppsala, Sweden
	}
 
   	\author{Olle Eriksson}
	\affiliation{
	Department of Physics and Astronomy, Uppsala University, Box 516, SE-75120 Uppsala, Sweden
	}
 	\affiliation{
	School of Science and Technology,  Örebro University, SE-701 82,  Örebro, Sweden
	}
 	\author{S\'ebastien Leb\`egue}
	\affiliation{
   Universit\'e de Lorraine, LPCT, CNRS, UMR 7019, BP 70239,  54506 Vandoeuvre-l\`es-Nancy Cedex, France
	}
	\author{St\'ephane Mangin}
	\affiliation{
	Universit\'e de Lorraine, IJL, CNRS, UMR 7198, BP 70239, 54000 Nancy Cedex, France
	}
	
	\date{\today}
 
	\begin{abstract}
Time resolved measurements of the linear magneto-optical Kerr rotation reveal that, in the {\it 3d} ferromagnets Fe, Co and Ni, the amplitude of the demagnetization increases linearly with the fluence of the light. We rationalize this phenomenon as being linearly driven by the increase in temperature, the electron-phonon and electron-magnon scattering, and a reduction of the inter-atomic exchange. The amplitude of the latter phenomenon, which until the present study has been widely overlooked, is obtained through \ai\ density functional theory calculations, and is argued to be the principal source of demagnetization in Ni and Co, while still contributing largely in Fe.
	\end{abstract}
	
	\pacs{Valid PACS appear here}
	\maketitle
	

More than twenty years ago, Beaurepaire \ea\cite{Beaurepaire1996} produced the first experimental evidence of the ultrafast quenching of the magnetic order by femtosecond light pulses. This discovery, and the perspective it held to allow for the manipulation of magnetization in unprecedented timescale, and without the use of a magnetic field, triggered an intense and very fruitful research that is still ongoing. Notably, a few years later, the all--optical switching, in which femtosecond light pulses allow for the deterministic manipulation of the magnetic state of metallic thin films was evidenced\cite{Stanciu2007, Radu2011a, Mangin2014, Lambert2014}. However, and regarless of their apparent complexity, the common denominator of these new types of magnetization dynamics, is the presence of an initial ultrafast demagnetization.
	
Naggingly, up to this day, the inner workings of this overarching phenomenon remain heavily debated, as much from the standpoint of the magnetic excitations responsible for the quenching, as regarding the types of mechanisms by which they are triggered. Early photoemission results\cite{Rhie2003, Cinchetti2006, Tengdin2018, You2018} pointed toward a reduction of the exchange splitting, suggesting induced Stoner excitations, thus inducing a longitudinal decrease of the atomic magnetic moments\cite{Scheid2022}. More recently Turgut \ea\cite{Turgut2016} and Eich \ea\cite{Eich2017} explicitly addressed this issue and experimentally showed a predominance of band mirroring in Co, a consequence of the presence of transversal excitations. As opposed to a decrease of the magnitude of each individual atomic magnetic moment, the demagnetization would therefore mainly result from a randomization of their direction. A similar conclusion has been drawn for bcc Fe\cite{Jana2020}.

To rationalize such dynamics, many theories have been proposed throughout the years. As reviewed by Scheid \ea\cite{Scheid2022}, possible sources of longitudinal excitations have been proposed; such as Elliot--Yafet electron-phonon spin-flip scatterings\cite{Carva2013}, a spin-dependent superdiffusive propagation of the hot-electrons\cite{Battiato2010}, a light--induced increase of the electronic temperature\cite{Scheid2019}, and a direct light--matter interaction.\cite{Krieger2015, Scheid2019a, Scheid2021} However, these theories were not able to explain the magnitude of the experimentally seen demagnetization, at least within the confinement of the experimentally absorbed energy.
On the other hand, ever since the pioneering work of Beaurepaire \ea\cite{Beaurepaire1996}, various temperature models have been a mean of understanding the characteristics of the demagnetization through a rapid increase of the temperature experienced by the magnetic moments\cite{Koopmans2010, Roth2012, Zahn2021, Pankratova2022}.
While such a framework does not explain how the angular momentum is conserved during the demagnetization, it provides the most straightforward rationalization of the latter, as originating from an ultrafast heating. Accordingly, in this regime, the demagnetization would occur through the generation of transversal excitations, or magnons, as they are the primary magnetic excitations responsible for the ferromagnetic-paramagnetic phase transition occuring at the Curie temperature in the elements studied here (\textit{e.g.} as pointed out by Gunnarsson\cite{Gunnarsson1976}).

In the present combined experimental and theoretical work, we demonstrate that the maximal demagnetization amplitude in the elemental transition metal (TM) ferromagnets Fe, Co and Ni challenges the aforementioned heating hypothesis. We argue that only accounting for an ultrafast heating of the magnetic degrees of freedom, the magnitude of which is regulated by electron-phonon and electron-magnon couplings as well as the incident light, hardly explains the experimentally seen demagnetization amplitude across the compounds.
Here we provide significant evidences that considering an overlooked phenomenon; the light-induced reduction of the inter-atomic exchange as an additional origin for the reduction of the magnetic order, resolves this conundrum. Indeed, while the impact of the light on the inter-atomic exchange has already been reported in iron oxide\cite{Mikhaylovskiy2015}, it has been overlooked in metals and even more as a possible origin for the ultrafast demagnetization. Within such a framework, rather than driving the magnetic system toward the Curie temperature (\Tc) through heating as explained within temperature models, \Tc\ is suggested here to be transiently decreased.
	
Glass/Ta(2)/TM(12)/Pt(2) multilayers were grown by RF (TM) and DC (Ta, Pt)-magnetron sputtering in a deposition tool with base pressure less than 3$\cdot 10^{-8}$ mbar. Float glass was used for the pump/probe geometry, the Ta(2) buffer layer promotes adhesion of the layers and block any oxygen diffusion, and Pt(2) capping layer is used to prevent layer oxidation.  The thickness of the films is large enough such that bulk properties are recovered, while the effect of absorption gradients are mitigated by the rapid spreading of hot electrons. The dynamics were probed using a magneto-optical microscope that monitors the longitudinal Kerr effect for spatially uniform s-polarized probe pulses having a wavelength of $515\,$nm and a duration of $150\,$fs as function of pump-probe delay. The latter were incident on the samples at an angle of $45\degree$. The normally incident, linearly polarized $800\,$nm, $150\,$fs pump pulses were focused on the sample to a $95\,\mu$m (FWHM) Gaussian. The pump was aligned to the center of the $560 \times 360\,\mu$m area of the sample displayed on the camera. The spatial resolution of the image was $\approx 10\,\mu$m. In-plane magnetic saturation fields were applied during the measurements to ensure full restoration of the magnetization between the pump-probe pulse pairs. The pulse repetition rate was 50 kHz and the exposure time of the camera was of the order of 1s, so that each image represents the average signal resulting from 50 000 laser pulses. The fluence dependence of the pump induced magnetization dynamics is derived from the measured images in three steps: first, potential artifacts (e.g. from pump induced changes of the non-magnetic sample reflectivity) are suppressed by subtracting the images measured for opposite magnetization directions at identical pump-probe delay, and dividing this difference by the one obtained at negative delay. Second, the spatial variations of the measured magnetization changes are mapped to the ones of the pump beam. Third, external fluences are converted to absorbed ones by means of the TMM python package and the complex indices extracted from Ref.\cite{rumble2022crc}.

\begin{figure}
\centering
\includegraphics[scale=0.8]{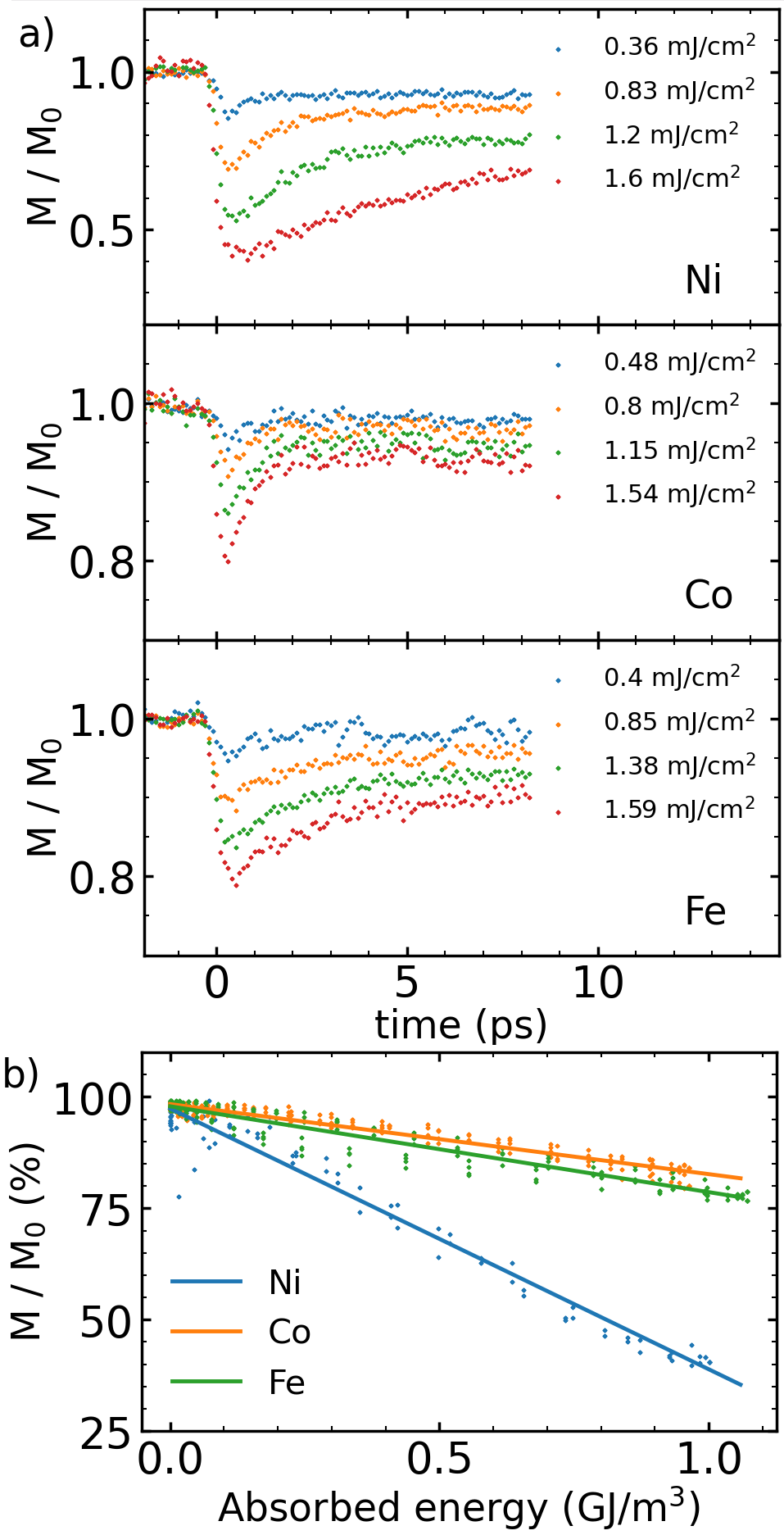}
\caption{a) Light--induced ultrafast demagnetization dynamics in Fe, Co and Ni at different absorbed fluences measured using the TR-MOKE. b) Normalized maximum demagnetization, m$_\text{min} = \frac{\text{M}_\text{min}}{\text{M}_\text{0}}$ as a function of the fluence. 
\label{fig:exp_dyn}}
\end{figure}

Fig. \ref{fig:exp_dyn} a) presents the magnetization dynamics for all three samples for different values of the absorbed fluence. Even though,  as expected, all the dynamics are characterized by a very swift reduction of the magnetization, these results challenge the existing viewpoints based on an ultrafast heating of the magnetic degrees of freedom on multiple accounts. To highlight this matter and provide a rightful comparison of the light-induced demagnetization amplitude of the different compounds, we utilize the fact that, just as \Tc\ does not depend on the saturation magnetization or, equivalently, the amplitude of the magnetic moment moments carried by each atom, but rather on the strength of their inter-atomic exchange interaction, so does the amplitude of the demagnetization. Indeed, the amplitude of the latter is expected to be only regulated by the ratio of heat reaching the bath of the magnetic moments on ultrafast timescales, and by the strength of the inter-atomic exchange. In Fig. \ref{fig:exp_dyn} b), we therefore consider the maximum relative demagnetization, m$_\text{min} = \frac{\text{M}_\text{min}}{\text{M}_\text{0}}$, where $M_0$ and $M_\text{min}$ respectively are the magnetization at saturation and the minimum magnetization reached during the dynamics. Doing so shows that the amplitude of the demagnetization increases linearly with the absorbed energy, \ie\ can be written as:
    
\begin{equation}
    	\text{m}_\text{min} = 1 + \alpha_\text{exp} \text{E}_\text{abs}
\end{equation}

where $\alpha_\text{exp}$ is the slope, which value is displayed in Tab.\ref{tab:recap} for the different compounds and $\text{E}_\text{abs}$ is the absorbed energy density, computed by dividing the absorbed fluences by the thicknesses of the absorbing films (16 nm).

Interestingly, this linear decrease of the magnetization with the absorbed energy contradicts the well established Bloch law stating that, upon thermal excitation, the variation of the magnetization is proportional to $T_m^{3/2}$, or equivalently $\propto E_m^{3/5}$ since $E_m \propto T_m^{5/2}$, where $T_m$ and $E_m$ respectively are the temperature and the energy of the magnons. Nonetheless, as suggested by the fact that it holds the  lowest \Tc\ (627 K), and therefore presents the weakest inter-atomic exchange interactions of the three compounds studied here, Ni is by far the most easily demagnetized element. On the other hand, and quite surprisingly such a reasoning fails when comparing Co, which has a \Tc\ of 1388 K, with Fe, that has a significantly lower \Tc\ of 1043 K, as both of them present very similar demagnetization amplitudes. 

In the framework of temperature models, this unexpected similarity in the value of $\alpha_\text{exp}$ for Co and Fe, could be the consequence of a larger electron-phonon coupling in Fe than in Co, thus more rapidly evacuating the absorbed energy from its electronic and spin degrees of freedom, and mitigating its demagnetization. Along this line of argument, the very pronounced demagnetization of Ni could originate from a very low electron-phonon coupling that would allow for the light--absorbed energy to almost solely affect the spin system and thus enhance the demagnetization. However, as can be seen in Tab.\ref{tab:recap} recapitulating \ai\ calculated values from Ritzmann \ea\cite{Ritzmann2020}, the electron-phonon coupling in Fe is more than three times weaker than it is in Co, while in Ni it is located in between, thus invalidating the aforementioned scenario.\\ 
Likewise, in a framework where the demagnetization is mostly due to transversal excitations, which, as discussed \textit{supra}, appears to be the case, a high electron-magnon coupling in Ni and Co as compared to Fe could explain the higher demagnetization in Ni and Co and the surprising resilience of Fe. However, as reported in Tab. \ref{tab:recap}, \ai\ calculations from Haag \ea\cite{Haag2014} in Fe and Ni show that the electron-magnon scattering rate in Ni is one order of magnitude lower than in Fe. While Haag \ea\cite{Haag2014} did not treat the case of Co, the work of Müller \ea\cite{Muller2019} indicates that it is similar to the one in Ni. Indeed, Müller \ea\ reported \ai\ calculations of the electronic lifetime broadening and band renormalization in Fe, Co and Ni as a consequence of the electron--magnon couling and found the strongest signatures of the latter in Fe, followed by Co and Ni. Consequently the electron-magnon coupling alone cannot explain the very large demagnetization in Ni and the resilience of Fe either.
 
Finally, in the framework of the microscopic three temperature model\cite{Koopmans2010, Roth2012}, the demagnetization rate is governed by the Elliot-Yaffet spin-flip scattering, and, if the electron-phonon coupling is low enough such that the electronic temperature stays high long enough, the initial demagnetization is followed by a remagnetization. Consequently, in this framework too, Fe, which has the lowest electron-phonon coupling and seemingly the largest Elliot-Yaffet spin-flip scattering rate of the three compounds studied here\cite{Carva2013}, should be more prone to feature a large demagnetization followed by a swift remagnetization occuring in less than 1 ps, the duration required for the equilibration of the temperature of the phonons and the electrons in Fe, as shown by Ritzmann \ea\cite{Ritzmann2020} than the other compounds. However, as seen in Fig. \ref{fig:exp_dyn}, the opposite is the case, \ie\ Fe features the slowest remagnetizing speed of the three compounds studied here.

\begin{table}
\centering
\begin{tabular}{l|c|c|c}
&  Ni    &  Co    &   Fe  \\
\hline 
Curie temperature (K)  &   627  &  1388  & 1043  \\
G$_\text{ep}$ (10$^{17}$W/(m$^3$K)) &  18.9  &  33.4  & 10.5  \\
W$^{\uparrow \downarrow}_\text{magnon}$  (scattering / 100 fs)  &  0.15  &  $\approx$ 0.15 & 1.34  \\
\hline
$\frac{dj}{dE}$ (\% GJ$^{-1}$ m$^3$)   & -48.40 & -15.82 & -7.85  \\
\hline
$\alpha_\text{exp}$ (\% GJ$^{-1}$ m$^3$)   & -58.51 & -15.71 & -19.27  \\
\end{tabular}
	
\caption{Summary of the Curie temperatures, \Tc, and the electron-phonon scattering rates, G$_\text{ep}$, from Ref. \cite{Ritzmann2020} and electrons--magnons scattering between two dominant spin states per 100 fs and per atom, $W^{\uparrow \downarrow}_\text{magnon}$ calculated at an electronic temperature of 900 K in Ref. \cite{Haag2014} for Ni and Fe. As discussed in the text, based on Ref. \cite{Muller2019} we assessed the value of $W^{\uparrow \downarrow}_\text{magnon}$ in Co to be close to the one in Ni. In addition, the slopes of the \ai\ calculated inter-atomic exchange reduction and of the experimental maximum demagnetization as a function as the absorbed energy are recapitulated.}
\label{tab:recap}
\end{table}

	
To sort out the apparent inconsistencies discussed above, we investigate the involvement of a previously overlooked phenomenon, which we find to be strongly correlated to the magnitude of the demagnetization: the ultrafast light-induced quenching of the inter-atomic exchange.
We use \ai\ density functional theory calculations\cite{Hohenberg1964, Kohn1965} to compute the dependence of the inter-atomic exchange on the light--absorbed energy in bcc Fe, fcc Co and fcc Ni. As in our previous work\cite{Scheid2019}, as well as in the framework of the temperature models, the increased electronic energy is accounted for as a rise of the electronic temperature, \Te, such that the occupation of the different Kohn--Sham states obey the Fermi-Dirac distribution. In this framework, the \Te-dependent magnons are obtained through a mapping of density functional theory spin-spiral energy calculations on the Heisenberg Hamiltonian\cite{Antropov1995, Antropov1996, Halilov1998}:
	
	\begin{equation}
		\hat{H}_i (t) = -\frac{1}{2}\sum_{i \neq j} J_{ij} \hat{\bm{M}}_j (t) \cdot \hat{\bm{M}}_i (t),
	\end{equation}

where $J_{ij}$ is the interaction energy between the two atomic sites $i$ and $j$.

To do so, we used the non-colinear full potential augmented planewave method (APW) and the LSDA\cite{Perdew1992} for the exchange correlation potential, as implemented in the ELK code. The APW basis is expanded in 10 spherical harmonics, $\left|k + G\right|_{max} R = 8$, where $R$ is the radius of the muffin–tin, and the Brillouin zone is sampled with a 28x28x28 mesh for all the compounds. A large grid is particularly important in order to insure the convergence at low \Te. The self–consistence is achieved when the total energy changes by less than 10\textsuperscript{-8} Ha in two successive electronic loops. The integrals over the first Brillouin zone required in the computation of the inter-atomic exchanges are performed on a 14$\times$14$\times$14 q-point sampling grid in a separate, home-made, python code.
As in these compounds, long range exchange interactions significantly contribute to the magnetic ordering\cite{Halilov1998, Pajda2001}, we  computed the effective value of the inter-atomic exchange, $j$, felt by an atomic site labeled by $\alpha$, by including all the interactions $J_{\alpha i}(E)$, comprised from the first, up to its 20\textsuperscript{th} neighbors as a function of the electronic energy $E(T_e)$:
	
\begin{equation}
		j(E) = \frac{\sum_i J_{\alpha i}(E)}{\sum_i J_{\alpha i}(E(T_e = 300 K))}
		\label{eq:J_eff}
\end{equation}

where the electronic energy, $E$, is readily extracted from the \Te\ dependent \ai\ calculations. Note that, as we are only interested in the rate of reduction of the inter-atomic exchange and not in its absolute value, in Eq. \ref{eq:J_eff}, we normalized by the same quantity computed at \Te\ $= 300$K.

\begin{figure}
\centering
\includegraphics[scale=0.25]{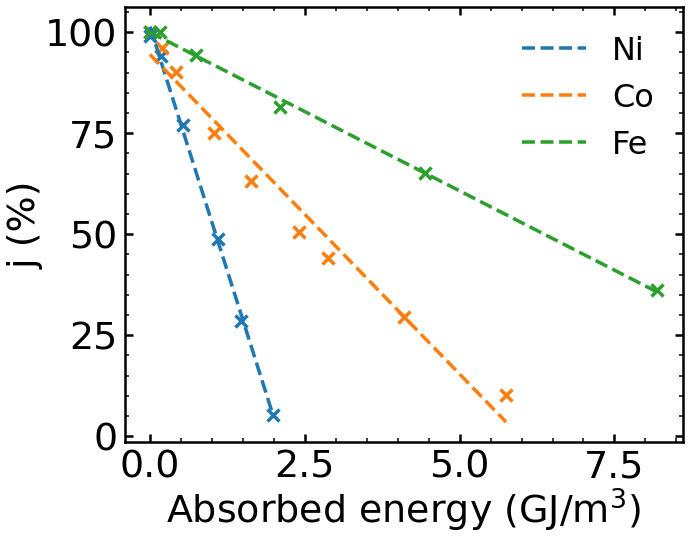}
\caption{Normalized inter-atomic exchange in Ni, Co and Fe, as a function of the electronic energy density obtained through \ai\ spin-spiral calculations. 
\label{fig:lin_j}}
\end{figure}

Fig. \ref{fig:lin_j}, featuring $j(E)$, shows that the inter-atomic exchange decreases linearly with the increased electronic energy. Quite remarkably, this reduction is far from being negligible in the range of the experimental fluences. This fact is better seen in Tab. \ref{tab:recap}, gathering the slopes of both the maximum demagnetization as a function of the absorbed energy density, shown in Fig. \ref{fig:exp_dyn} b), and the slope of the decrease of $j$ as a function of the absorbed energy shown in Fig. \ref{fig:lin_j}. Strikingly, in Ni and Co both rates of decrease are matching quite closely, while in Fe the rate of the decrease of the exchange is half the one of the maximum demagnetization.
This apparent correlation between the maximum demagnetization and the \ai\ calculated reduction of the exchange points toward a mechanism where, instead of, or on top of being the consequence of a rapid heating, the demagnetization could be due to a quenching of the interaction responsible for the magnetic ordering itself. Indeed, upon reduction of the inter-atomic exchange, the ferromagnetic ordering is expected to decrease through the creation of magnons. While the rate at which magnons are generated may be tied to the damping, the reduction of the inter-atomic exchange is due to its parametric dependence on the electronic state and thus occurs instantaneously with the perturbation brought by the light-pulse, thus instantaneously affecting the magnon states. Such a phenomenology is in direct contrast with a transfer of heat from the electrons to the magnons as assumed by temperature models and, given the fact that the inter-atomic exchange in Fe is far less affected by the absorption of the light than Co, could explain their similar demagnetization amplitude even though Fe features a much lower \Tc.

To investigate this possibility, we assume that the demagnetization proneness of the different compounds, characterized by $\alpha_\text{exp}$, scales linearly with the amplitude of the different phenomena taking place during the light-induced demagnetization. As usually described in temperature models, the latter include: a heating of the magnetic degrees of freedom, which is assumed to scale as the inverse of \Tc, the electron-phonon coupling ($G_{ep}$), the electron-magnon coupling ($G_{em}$) and the reduction of the inter-atomic exchange interaction, $\frac{dj^{\text{C}}}{dE}$. The resulting system of equation therefore writes as:

	\begin{equation}
		\begin{split} 
		\alpha^{\text{C}}_\text{fit}(x_{T},  x_{j},  x_{G_{ep}}, x_{G_{em}}) =\\
		 x_{T} \frac{1}{T_C^{\text{C}}} + x_{j} \frac{dj^{\text{C}}}{dE} + x_{G_{ep}} G_{ep}^{\text{C}} + x_{G_{em}} G_{em}^{\text{C}},
		\end{split}
		\label{eq:slope_fit}
	\end{equation}

with $x_{T}$, $x_{j}$, $x_{G_{ep}}$ and $x_{G_{em}}$ the fitting coefficients which are independent of the compound as characterize the involvement of the aforementioned phenomena in the demagnetization process, and $\alpha^{\text{C}}_\text{fit}$ the fitted slope for the different compounds indexed by $\text{C}=\left\{ \text{Fe}, \text{Co}, \text{Ni} \right\}$.

\begin{figure}
\centering

\includegraphics[scale=0.27]{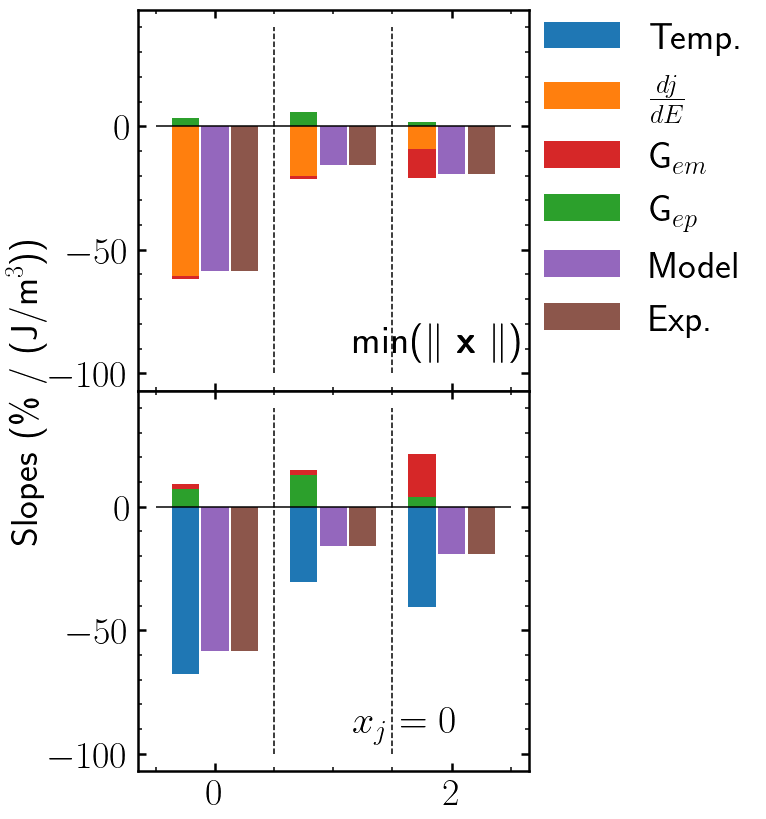}

\caption{Contribution to the demagnetization of: (1) thermal fluctuation, $x_{T} \frac{1}{T_C^{\text{C}}}$, in blue (2) reduction of $j$ as a function of the absorbed energy, $x_{j} \frac{dj^{\text{C}}}{dE}$, in orange (3) electron-phonon, $x_{G_{ep}} G_{ep}^{\text{C}}$, in green and (4) electron-magnon coupling, $x_{G_{em}} G_{em}^{\text{C}}$, in red to $\alpha_\text{fit}^\text{C}$, the fitted slope of the maximum demagnetization in purple for Ni, Co and Fe.
$\alpha_\text{fit}^\text{C}$ is computed in three different cases:  (Top) minimizing the norm of $\bm{x} = \left(x_{T}, x_{j}, x_{G_{ep}}, x_{G_{em}}\right)^\text{T}$, (Bottom) setting $x_j = 0$. For comparison the value of the experimental slope, $x^\text{C}_{\text{exp}}$ is displayed in brown.
\label{fig:slopes_contrib}}
\end{figure}
	
Fig. \ref{fig:slopes_contrib} shows $\alpha_\text{fit}^{\text{C}}$ (purple bar) as well as the different contributions to it (blue, orange, red and green), for Fe, Co and Ni. The figure also shows $\alpha^{\text{C}}_\text{exp}$ (brown) for all elements investigated here. As the system of equations is under-determined, \ie\ there are four variables, $\bm{x} = \left(x_{T}, x_{j}, x_{G_{ep}}, x_{G_{em}}\right)^\text{T}$, and only three constraints given by the experimental slopes of the three elements, we look for the solution minimizing $\| \bm{x} \|$. Doing so provides the simplest explanation of the experimental slopes. Interestingly, as shown in Fig. \ref{fig:slopes_contrib}, such an analysis completely excludes a contribution originating from a rise of the temperature as the explanation for the amplitude of the demagnetization and, instead, attributes it mainly to the quenching of the inter-atomic exchange in Ni and Co. This is well explained by the aforementioned very good agreement between $\frac{dj^{\text{C}}}{dE}$ and $\alpha_\text{exp}$ for these two compounds (see Tab. \ref{tab:recap}). On the other hand, in Fe, and in agreement with the work of Carpene \ea\cite{Carpene2008}, the electron--magnon scattering explains more than half of the demagnetization, while the reduction of the inter-atomic exchange explains the remaining contribution. Supposing that the ``quick recovery" regime\cite{You2018} is due to a recovery of the inter-atomic exchange, this may explain why Fe does not feature this phenomenon as strongly as Ni and Co do. Here we also notice that the model correctly attributes a positive sign to $x_{G_{ep}}$, which is in accordance with the aforementioned physical insight that the electron-phonon coupling tends to reduce the amplitude of the maximum demagnetization.\\
The case when $x_j$ is imposed to be zero is also considered here, \ie\ placing ourselves in the usual framework in which the reduction of exchange is neglected. In its absence, and in accordance with the introductory discussion about the impossibility for current theoretical frameworks to describe these experimental data, the model provides inconsistent results: the contribution of the electron-magnon scattering is positive, \ie\ tends to limit the amplitude of the demagnetization. This emphasizes the necessity of accounting for the involvement of the reduction of the inter-atomic exchange to correctly describe the demagnetization amplitude across the studied compounds.\\

To conclude, the reduction of the inter-atomic exchange provides a novel mechanism through which transversal excitations, or magnons can be generated on the ultrafast timescales. Indeed, rather than destabilizing the magnetic ordering by a straightforward light-induced heating, the initial part of the demagnetization can, in a significant part, be due to a transient reduction of the inter-atomic exchange responsible for the ferromagnetic ordering itself. This work is therefore in line with recent experimental measurements attributing the main part of the demagnetization as resulting from transversal excitations\cite{Turgut2016,Eich2017,Scheid2022}. Moreover, in agreement with the experimental work of Bergeard \ea\cite{Bergeard2016} this mechanism does not rely on a direct light--matter interaction, but rather on the perturbation of the inter-atomic exchange due to the presence of hot electrons.
Finally, considering the reduction of the inter-atomic exchange as an additional source of demagnetization provides an explanation for the paradoxically similar demagnetization amplitude experimentally observed in Fe and Co. As this phenomenon is particularly strong in Ni, it could be responsible for the critical behavior occurring within 20 fs of the ultrafast demagnetization evidenced by Tengdin \ea\cite{Tengdin2018}.\\

\begin{acknowledgments}
This work is supported by the ANR-20-CE09-0013 UFO. O.E acknowledges support from the Swedish Research Council. O.E. and A.B. acknowledges support from the Knut and Alice Wallenberg Foundation. P. Scheid is grateful for the many fruitful discussions with Q. Remy, A. Szilva, I. P. Miranda and M. Pankratova.
\end{acknowledgments}

\end{document}